\documentstyle[12pt]{article}

\begin{document}
\input epsf.tex

\rightline{McGill/96-13}
\vspace{.5 cm}
\begin{center}
{\Large\bf Cut Diagrams for High Energy Scatterings}\\
\vspace{.5 cm}
{Y.J. Feng$^*$, O. Hamidi-Ravari$^{\dag}$ and C. S. Lam$^{\ddag}$}\\
\bigskip
{\it Department of Physics, McGill University,\\
3600 University St., Montreal, P.Q., Canada H3A 2T8}
\end{center}

\begin{abstract}
A new approach is introduced to study QCD amplitudes at high energy and
comparatively small momentum transfer. Novel cut diagrams,
representing resummation of Feynman diagrams, are used to simplify
calculation and to avoid delicate cancellations encountered in the usual
approach. Explicit calculation to the 6th order is carried out to
demonstrate the
advantage of cut diagrams over Feynman diagrams.
\end{abstract}

\section{Introduction}
The rapidity-gap events recently observed at HERA \cite{zu} revived the community's
interest in Regge-pole description of high energy scattering.
The word `Pomeron' seldom heard in recent years is once again found
in the lexicon of experimentalists. It is perhaps then a good time
to have a new look at the connection between perturbative QCD and Regge pole.
Non-perturbative effect may be important for
momentum transfer of the order of $\Lambda_{QCD}$ or smaller, but we shall
avoid it by going to a larger momentum transfer if necessary.

It was proposed by Low and Nussinov \cite{lo} some years ago that the Pomeron may
simply reflect a two-gluon exchange in QCD. Diagrammatic calculations
have been carried out
\cite{CT,NT,LN,JB,LH,EA} in the leading log approximation  to substantiate this proposal and
to study other aspects of the high energy near forward scattering amplitudes.
We shall follow the classical approach of implementing physical ($s$-channel)
unitarity by summing up Feynman diagrams in the Feynman gauge \cite{CT,NT}. However,
other approaches are available. The use of physical gauge and dispersion
relation \cite{LN} can sometimes make things more transparent. One can also
emphasize on the complex angular momentum aspect by concentrating on
$t$-channel towers and their interactions as guided by $t$-channel unitarity
\cite{JB}. We shall speak no further of these alternate approaches except to remark
that our cut diagram method, to be discussed below, can be thought of
an intermediate link between the classical Feynman diagram sum and the
dispersion relation approaches.

Due to the gauge and non-abelian nature of
the theory, perturbative calculations are lengthy and complicated. Even by
ignoring self energy and vertex corrections, as well as renormalization effects
on the grounds that they will not alter the qualitative
nature of high energy scattering we are trying to learn,
even by ignoring quark pair productions as a first step,
a complete QCD calculation can be carried out only up to the 6th order.
In the much simpler case of QED, an 8th order calculation
\cite{HC} was reported to have taken sixteen
months and two thousand pages to complete.  In the case of QCD, it is much
worse, a complete
calculation of the 8th order is not available, though partial calculations
have been carried out \cite{LH}.

What makes the calculation so complicated is the large number of diagrams
that has to be tackled, and the inevitable cancellations between them. 
The computation of each diagram is fairly straight
forward, though somewhat lengthy at higher orders. After the individual
diagrams are calculated, care must also be exercised to add them up because of
the presence of many delicate cancellations. Leading-log dependences
on energy get subtracted away, complicated functions of momentum transfer
also disappear. What emerges at the end is a product that is surprisingly
simple. To the extent that it has been verified, high energy near forward
scattering is described by multiple reggeized gluon exchange, supplemented
by elementary gluon production off the reggeons and by $s$-channel unitarity
\cite{CT,LH}.

This outcome has been verified up to the 6th order \cite{CT,NT,LN}.
As mentioned above, many delicate cancellations take place in the sum to
enable this simple picture to emerge. A complete calculation in the 8th order
is not available, but if one assumes these cancellations which take place
up to the 6th order also occur in the 8th and higher orders, then this
reggeized picture has been verified up to the 10th order\cite{LH}. This gives
a strong support to the conjecture that it is true to all perturbative orders.

While the conjecture is attractive, it is impossible to verify or refute
without a new method to simplify the complicated calculations.
We report in this article such a new method, in which high energy scatterings
are computed via {\it cut diagrams} rather than the normal Feynman diagrams.
These cut diagrams are {\it not} the Cutkosky cut diagrams. The sum of all
cut diagrams here is equal to the sum of all Feynman diagrams, and not just
their discontinuities. These cut diagrams can be regarded as a resummation
of the Feynman diagrams
in which many of the delicate cancellations encountered in the latter
have been built in and explicitly avoided. Consequently, the simplicity of the
final sum
is revealed already in individual cut diagrams, and not masked by terms to be
cancelled as is the case with Feynman diagrams. Besides, individual cut
diagrams are easier to calculate than individual Feynman diagrams.

In this paper we shall introduce the formalism of cut diagrams, as well
as an explict calculation to the 6th order to demonstrate its effectiveness.
We shall discuss quark-quark scattering throughout but allow the quarks
to carry any $SU(N)$ color charge. The result should be equally valid
for gluon-gluon and other scatterings because high energy processes are
insensitive to the spins of the colliding particles. We will study
pure $SU(N)$ QCD and ignore the production of quark pairs.
If one is bothered by considering quark-quark scattering while ignoring
quark productions, one can consider instead gluon-gluon scattering, and
as mentioned above, the result would be the same.

In a subsequent paper\cite{YJ} we shall use the cut diagrams to discuss
the reggeization conjecture in higher orders.

In Sec.~2 the result of Feynman diagram calculations to $O(g^6)$ is
reviewed.
How these diagrams combine to give a sum much simpler than the individuals
will be discussed in some detail. In Sec.~3, we begin to examine more
carefully these cancellations by using
sum rules. In Sec,~4, the method of cut diagrams is explained, and in
Sec.~5, this method is applied to 4th and 6th order calculations to
demonstrate the savings effected by this method.

\section{ Elastic scattering up to $O(g^6)$}

Leading-log calculations in QCD,
at high c.m.~energy $\sqrt{s}$
and comparatively small momentum transfer $\sqrt{-t}=\sqrt{\Delta^2}$,
are discussed, among others\cite{BW}, in the book of Cheng and Wu \cite{CT}. Citations to
the original literature can also be located there. We review in this section
the calculation of quark-quark scattering to the 6th order done in this
standard way.

\begin{figure}
\vskip -0 cm
\centerline{\epsfxsize 3 truein \epsfbox {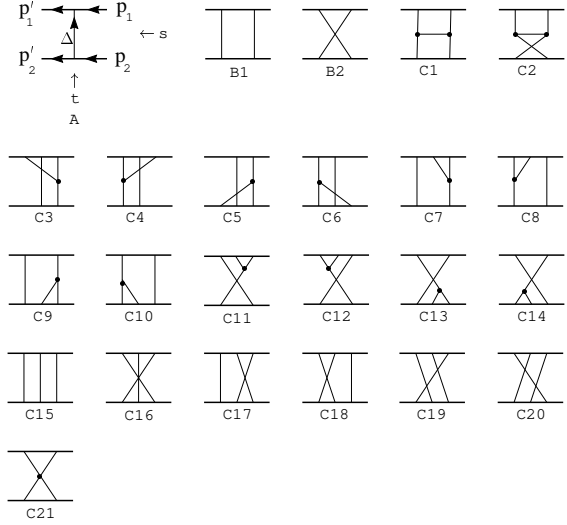}}
\nobreak
\vskip -2 cm\nobreak
\vskip .1 cm
\caption{Quark-quark scattering in QCD up to the 6th order. The thick lines at the top and bottom of each diagram are the fermion lines, and the thin lines are gluon lines. }
\end{figure}

The relevant Feynman diagrams are given in Fig.~1. The 2nd order
diagram is labelled A, the 4th order diagrams  labelled B1 and B2, and the
6th order diagrams  C1 to C21. This last labelling
is identical to the ones used in Fig.~12.7 of Ref.~\cite{CT}\cite{note1}.

Under the interchange of the
Mandelstam variables $s=(p_1+p_2)^2$ and $u=(p_1-p_2')^2$,
the spacetime parts of these diagrams are either self conjugate, or
one
goes into another. For example,
A${\leftrightarrow}$A, B2${\leftrightarrow}$B1, C2${\leftrightarrow}$C1,
C15${\leftrightarrow}$C16, C17${\leftrightarrow}$C19, and C18${\leftrightarrow}$C20, under $s{\leftrightarrow} u$.

Normalizing the Dirac spinors to $\overline{u} u=1$, and designating the
fermion mass and gauge coupling constant as $m$ and $g$, with $\beta=g^2/2\pi$,
the T-matrix element
${\cal T}=(g^2s/2m^2){\cal M}$ receives the following contributions from the individual
diagrams to ${\cal M}$ [3,4]\cite{note2}:
\begin{eqnarray}
A&=&-I_1\!\cdot\!{\bf G}_1\nonumber\\
B1&=&- \beta\ln\left(  se^{-\pi i}\right) I_2\!\cdot\!{\bf G}_2\nonumber\\
B2&=&+\beta(\ln s) I_2\!\cdot\!({\bf G}_2+c{\bf G}_1)\nonumber\\
\overline{C1}&=&+\beta^2\ln^2\left(se^{-\pi i}\right)\left[ {1\over 2}
\Delta^2I_2^2-J_2I_2\right] \!\cdot\!{\bf G}_3\nonumber\\
\overline{C2}&=&-\beta^2({\ln^2s})\left[ {1\over 2}
\Delta^2I_2^2-J_2I_2\right] \!\cdot\!({\bf G}_3+c^2{\bf G}_1)\nonumber\\
C3&=&+\beta^2\{\ln^2\left(se^{-\pi i}\right)-{\ln^2s}\}
{1\over 4}J_2I_2\!\cdot\!({\bf G}_3-c{\bf G}_2)\nonumber\\
&=&C4=C5=C6\nonumber\\
C7&=&-\beta^2\ln^2\left(se^{-\pi i}\right){1\over 4}J_2I_2\!\cdot\!(-c{\bf G}_2)\nonumber\\
&=&C8=C9=C10\nonumber\\
C11&=&+\beta^2({\ln^2s}){1\over
4}J_2I_2\!\cdot\!(-c{\bf G}_2-c^2{\bf G}_1)\nonumber\\
&=&C12=C13=C14\nonumber\\
C15&=&-\beta^2(\ln s) 2J_3\!\cdot\!{\bf G}_4\nonumber\\
C16&=&-\beta^2(\ln s) 2J_3\!\cdot\!({\bf G}_4-{\bf G}_3+3c{\bf G}_2+c^2{\bf G}_1)\nonumber\\
C17&=&+\beta^2(\ln s)(J_3+\pi iI_3)\!\cdot\!({\bf G}_4+c{\bf G}_2)=C18\nonumber\\
C19&=&+\beta^2(\ln s)(J_3-\pi
iI_3)\!\cdot\!({\bf G}_4-{\bf G}_3+2c{\bf G}_2)=C20\ .
\end{eqnarray}
Part of $C21$ has been combined with $C1$ to form $\overline{C1}$, and the
remaining part of $C21$ has been combined with $C2$ to form $\overline{C2}$.
Note that the result for different diagrams is consistent with
their $s{\leftrightarrow} u$ character, under whose exchange the ${\cal T}$ amplitude
responds with the swap $se^{-\pi i}{\leftrightarrow} s$. See App.~A and B for a brief
discussion of how these amplitudes are computed.

The two-dimensional vector $\Delta$ in the transverse direction is the
momentum transfer. The functions $J_n(\Delta)$ and $I_n(\Delta)$ are
defined as follows:
\begin{eqnarray}
I_1(\Delta)&=&{1\over \Delta^2}\nonumber\\
I_n(\Delta)&=&\int \left(  \prod_{i=1}^n {d^2q_{i\perp}\over(2\pi)^2}{1\over
q_{i\perp}^2}\right)(2\pi)^2\delta^2\left(  \sum_{i=1}^nq_{i\perp}-\Delta\right)\nonumber\\
J_2(\Delta)&=&\int  {d^2q_{\perp}\over(2\pi)^2}{1\over
q_{\perp}^2}\nonumber\\
J_3(\Delta)&=&
\int \left(  \prod_{i=1}^3 {d^2q_{i\perp}\over(2\pi)^2}{1\over
q_{i\perp}^2}\right)\ln{q_{2\perp}^2\over q_{3\perp}^2}
\cdot(2\pi)^2\delta^2\left(  \sum_{i=1}^n{q_{i\perp}}-{\Delta}\right)
\ .
\end{eqnarray}
The functions $I_2, I_3, J_2, J_3$ are denoted respectively  by $I,
I_1, K, I_2$  in Ref.~\cite{CT}.
The infrared divergences of these integrals can be
regulated by a mass, either put in by hand or via the Higgs mechanism.
This regulation discussed in the
literature [3--9] does not affect the following
discussions so we shall ignore it.
There is however also an ultraviolet divergence in the
integral defining $J_2(\Delta)$, but it turns out that this function disappears
in the {\it sum} of the sixth order diagrams so it causes no trouble either.

The factors ${\bf G}_i$ in (2.1) are the color factors, with $c=N/2$ for $SU(N)$
colors.  As defined in \cite{CT}, ${\bf G}_1,{\bf G}_2,{\bf G}_3,{\bf G}_4$ are respectively the color
factors for the diagrams A, B1, C1, and C15. They can be computed by
an elegant graphical method \cite{CT}
from the $SU(N)$ commutation relation and identities

\begin{eqnarray}
[t_a,t_b]=if_{abc}t_c\ ,\quad
f_{abc}f_{abd}=2c\delta_{cd}\ ,
\quad
i^3f_{adg}f_{bed}f_{cge}=cif_{abc}\ .
\end{eqnarray}
Hence the combination of color factors ${\bf G}_i$ given in (2.1) for the
various diagrams remain valid whatever the color of the
quark is, although
${\bf G}_i$ themselves would be different for different color of the quark.
See App.~A for a brief discussion on the computation of the color factors.

The sum of all the terms in (2.1), from $A$ to $C20$, is
\begin{eqnarray}
{\cal M}&=&-{1\over\Delta^2}\left[ 1-\overline{\alpha} \ln s+{1\over 2}\overline{\alpha} ^2{\ln^2s}\right] \!\cdot\!{\bf G}_1
+{1\over 2} g^2i(I_2-2\beta cI_3\ln s)\!\cdot\!{\bf G}_2\nonumber\\
&+&g^2i\beta\ln s\left[ I_3-{1\over 2}\Delta^2I_2^2\right] \!\cdot\! {\bf G}_3
+g^4{1\over 6}I_3\!\cdot\! {\bf G}_4\ ,
\end{eqnarray}

with
\begin{eqnarray}
\overline{\alpha} (\Delta)\equiv \beta c\Delta^2I_2(\Delta)\ .
\end{eqnarray}

It is important to note the various cancellations taking place
to make the sum (2.4)
vastly simpler than the individual terms appearing in (2.1).
For example,
\begin{enumerate}
\item
In the fourth order, the leading term proportional to $\ln s$ is cancelled
out between $B1$ and $B2$ in the color amplitude proportional to ${\bf G}_2$,
though not in ${\bf G}_1$.
\item In the sixth order, the leading $\ln s$ contributions to ${\bf G}_4$
from $C15$ to $C20$ also add up to zero. The expressions given in (2.1)
are not accurate enough to deal with the subleading terms.
The term in (2.4) proportional to ${\bf G}_4$ is obtained separately from
the eikonal formula.
\item As a result of these cancellations,
the energy dependence and the $SU(N)$ (or $c$) dependence of the ${\bf G}_1$
amplitude is $(g^2c\ln s)^m$, and
those of ${\bf G}_2, {\bf G}_3, {\bf G}_4$ are respectively $g^2(g^2c\ln s)^m$, $g^2(g^2\ln s)^m$,
$g^4(g^2c\ln s)^m$. These dependences can be summarized all at once by
introducing
a different notation for the color factors. We shall use the notation
${\bf F}_{i,j}$ to denote a color factor with $i$ parallel vertical
lines connecting the two fermions, and $j$ parallel horizontal
lines joining any two of the vertical gluon lines. We shall also write
${\bf F}_{i,0}$
simply as ${\bf F}_i$. The relations with the color factors ${\bf G}_i$ are ${\bf G}_1={\bf F}_1,
{\bf G}_2={\bf F}_2, {\bf G}_3={\bf F}_{2,1}$, and ${\bf G}_4={\bf F}_3$.
The $g, c$ and $\ln s$ dependences of ${\bf F}_{i,j}$ in (2.4) are then given by
$g^{2(i-1)}(g^2c\ln s)^m c^{-j}$ for a diagram of order $2(m+i)$.
We shall refer to such dependences as {\it Regge-like}, for the Reggeization
of the scattering amplitude to be discussed later relies critically on this
feature of the scattering amplitude. Note from (2.1) that contributions
from individual diagrams are not Regge-like. Only the sum is.
\item
Simplification in transverse-momentum dependences also occur in the sum.
The simple integrals $I_n$ survive, but
the complicated integral $J_3$ and the divergent integral $J_2$
do not appear in the sum. This cancellation is highly nontrivial because both
of them contribute
different amounts to different color amplitudes. More specifically,
\item The function $J_3(\Delta)$
appears in all the color amplitudes ${\bf G}_1,{\bf G}_2,{\bf G}_3$ and ${\bf G}_4$ in diagrams
C15 to C20. Those in ${\bf G}_2,{\bf G}_3,{\bf G}_4$ actually get cancelled out in the sum,
but its presence in the ${\bf G}_1$ amplitude survives. However,
since this term is of order $g^6\ln s$, it is negligible compared to terms
of order $g^6{\ln^2s}$ appearing in the ${\bf G}_1$ amplitudes of $\overline{C2}$
and $C11$ to $C14$, it can be ignored in the leading-log result
displayed in (2.4).
\item
$J_2(\Delta)$ appears in
the color amplitudes ${\bf G}_1,{\bf G}_2$ and ${\bf G}_3$ in individual diagrams C1 to C14
and all these appearances get cancelled out.

\end{enumerate}
As a result of these cancellations, ${\cal M}$ acquires a very simple
interpretaion in terms of reggeized gluon exchanges .
These exchanges are constructed in such a way to ensure $s$-channel unitarity
[3,7].

Let us denote the reggeon propagator by
\begin{eqnarray}
R_1(\Delta,s)={1\over \Delta^2}\exp(-\overline{\alpha} (\Delta)\ln s)\ .
\end{eqnarray}
This reduces to the (transverse part of the) ordinary
propagator $I_1(\Delta)=\Delta^{-2}$ for small $g^2c\ln s$. Similarly, let us denote the
reggeized version of $I_n(\Delta)$ by
\begin{eqnarray}
R_n(\Delta,s)=\int \left(  \prod_{i=1}^n {d^2q_{i\perp}\over(2\pi)^2}
R_1(q_{i\perp},s)\right)\cdot(2\pi)^2\delta^2\left(  \sum_{i=1}^nq_{i\perp}-\Delta\right)\ ,
\end{eqnarray}
indicating the exchange of $n$ reggeons. Then to order $g^6$ in ${\cal T}$,
we can write
\begin{eqnarray}
{\cal M}&=&-R_1(\Delta,s)\!\cdot\!{\bf F}_1+ i{g^2\over 2!}\left[ R_2(\Delta,s)\!\cdot\!
{\bf F}_2+R_{2,1}(\Delta,s)\!\cdot\!{\bf F}_{2,1}\right]  \nonumber\\
&+&{g^4\over 3!}R_3(\Delta,s)\!\cdot\!{\bf F}_3\ .
\end{eqnarray}
In other words, the ${\bf F}_1,{\bf F}_2, {\bf F}_{2,1}$, and ${\bf F}_3$ components looked
precisely like diagrams A, B1, C1, and C15 respectively, but with the
vertical gluons replaced by their reggeized version whose propagators are given
in (2.6), {\it and} with all longitudinal-momentum integrations omitted.
To interpret it this way for $R_{2,1}{\bf F}_{2,1}$ we need to know the
Lipatov-Dickinson vertex \cite{EA} describing how elementary gluons are produced
and absorbed from the reggeized gluons.

This remarkable simplicity and regularity led to the conjecture \cite{CT,LH} that
the reggeized formula (2.8), suitably generalized, is the correct high energy
limit to all
perturbative orders. This conjecture is very difficult to verify on account of
the shear complexity in higher order calculations.
For QCD in the 8th order
it is simply not manageable without simplifying assumptions. If one assumes
all cancellations occured up to $O(g^6)$ will also
occur in higher orders, the final result can be extracted
from a {\it relatively} small set of diagrams, then it is reported that this
reggeization conjecture is true to 8th and 10th orders \cite{LH}.
Even so these heroic calculations are so lengthy and complicated that
to our knowledge the full details have never been published.
\setcounter{equation}{0}
\section{High energy kinematics}

We shall discuss in the next two sections a method of using {\it cut diagrams}
to sum up the Feynman diagrams,
a method in which most of the cancellations discussed in the last section are
automatically built in. This simplification shortens the computations
and makes it possible to study higher order
diagrams. In this paper we shall discuss how it simplifies
the calculations up to $O(g^6)$. In a subsequent paper \cite{YJ} we shall discuss
how it helps to verify part of the reggeization conjecture to all orders.
To prepare for the grounds for both, we discuss here the
relevant kinematical features of high energy scattering which enables this
new method to work.

We will assume the colliding beams in their c.m.~system to be directed along
the
z direction. In lightcone coordinates, $p_\pm=p^0\pm p^3$, the components of
a four-vector are labelled in the order $p^\mu=(p_+,p_-,p_\perp)$,
with the two-dimensional vector $p_\perp$ lying in the transverse x--y plane.
In this notation, the incoming fermion momenta are $p_1=(\sqrt{s},0,0)$ and
$p_2=(0,\sqrt{s},0)$, in which their mass $m$ has been neglected.
The outgoing fermion momenta are approximately given by
$p_1'=(\sqrt{s},0,\Delta)$ and $p_2'=(0,\sqrt{s},-\Delta)$. See Fig.~1.

Suppose $n$ gluons are hooked up to the upper fermion line as shown
in Fig.~2. The initial and final fermions are on-shell
but the gluons can be off-shell, though with an amount of energy far less than
$\sqrt{s}$.  At high
energy,
the numerator of the propagator can be approximated by
\begin{eqnarray}
\gamma p=2m \sum_\lambda u_\lambda(p_1)\bar u_\lambda(p_1)
\end{eqnarray}
provided the Dirac spinors are normalized to 
$\bar u_\lambda(p)u_{\lambda'}(p)=\delta_{\lambda\lambda'}$.
With that, the dominant current 
$\bar u_\lambda(p_1)\gamma^\alpha
u_{\lambda'}(p_1)$ at high energy is just its translational part
$\delta_{\lambda\lambda'}p_1^\alpha/m$. This shows that the spin content at
high energy is unimportant. All that it does is to enforce helicity
conservation of the fermion, and to produce a factor
$2p_1$ at each vertex together with an overall normalization factor of
$1/2m$. For most of the discussions below we shall ignore this QED factor and
concentrate  on the contribution from the denominators, {\it i.e.,} the
corresponding scalar theory.

\begin{figure}
\vskip -.2 cm
\centerline{\epsfxsize 3.2 truein \epsfbox {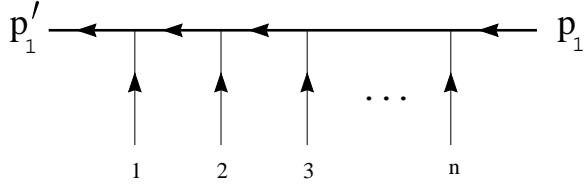}}
\nobreak
\vskip 0 cm\nobreak
\vskip .1 cm
\caption{ A quark $\to$ quark + $n$ gluon tree diagram. }
\end{figure}

The denominator of the $i$th inverse propagator is
\begin{eqnarray}
(p_1+\sum_{j=1}^iq_j)^2-m^2+i\epsilon\simeq
s(\sum_{j=1}^ix_j+i\epsilon)\ 
\end{eqnarray}
where $x_i=q_{i-}/\sqrt{s}$.
The scalar amplitude in Fig.~2 is given by $s^{-n}$ times
\begin{eqnarray}
a[12\cdots n]\equiv -2\pi
i\delta\left(  \sum_{j=1}^nx_j\right)\prod_{i=1}^{n-1}{1\over\sum_{j=1}^ix_j+i
\epsilon}\ .
\end{eqnarray}
Note that a momentum conservation $\delta$-function for the negative components
(together with an explicit factor $-2\pi i$) has been incorporated.
In eq.~(3.3) the ordering of the gluon lines from left to right is $[123\cdots n]$.
If they are ordered differently, say $[v_1v_2\cdots v_n]\equiv V$, then the
corresponding amplitude is
\begin{eqnarray}
a[v_1v_2\cdots v_n]\equiv a[V]
\equiv -2\pi
i\delta\left(  
\sum_{j=1}^nx_{v_j}\right)\cdot\prod_{i=1}^{n-1}{1\over\sum_{j=1}^ix_{
v_j} + i \epsilon}\ .
\end{eqnarray}
\setcounter{equation}{0}
\section{Sum Rules}

It is possible to compute sums of Feynman diagrams without much of
the delicate cancellations discussed in Sec.~2, by using the
{\it cut diagrams} which we shall describe in this section and the next.
The derivation of cut diagrams relies on two exact combinatorial formulas
for the quantity $a[V]$ in (3.4):
the {\it factorization formula} and the
{\it multiple commutator formula} derived in Ref.~\cite{CS}. We shall discuss the
former in this section, and the latter in the next section.

Consider an ordering of $n_i$ gluon lines:
$[v_{i1}v_{i2}\cdots v_{in_i}]\equiv V_i$. We shall use the notation
$\{V_1;V_2;\cdots;V_m\}$ to denote the {\it set} of {\it all} orderings
of
the $M\equiv\sum_{i=1}^mn_i$ gluon lines, {\it provided} the relative orderings
of
lines within each $V_i$ are maintained. The number of orderings in
this set is given by the multinominal coefficient $M!/\prod_{j=1}^mn_j!$.
For example, if $V_1=[135], V_2=[24]$,
then $\{V_1;V_2\}\equiv\{135;24\}$ consists of the $5!/3!2!=10$ orderings
[13524], [13254], [13245], [12354], [12345], [12435], [21354], [21345],
[21435], and $[24135]$.

We shall use the notation
\begin{eqnarray}
a\{V_1;V_2;\cdots ;V_m\}=\sum_{V\in\{V_1;V_2;\cdots ;V_m\}}a[V]
\end{eqnarray}
to denote the sum of
all amplitudes for the gluon orderings in the set.
The factorization formula \cite{CS} then states that
\begin{eqnarray}
a\{V_1;V_2;\cdots ,V_m\}=\prod_{i=1}^ma[V_i]\ 
\end{eqnarray}
In particular, if each set $V_i=[v_i]$ consists of only one gluon line labelled
by $v_i$, then $\{V_1;V_2;\cdots ;V_m\}$ is the set of {\it all} orderings
of the $m$ gluon lines. In that case the factorization formula reduces to
the well-known {\it eikonal formula} \cite{CT,HC1}. Other special cases of this
formula have also been discovered before\cite{LH,CY}.

It is useful to adopt an alternative notation for the right hand side of (4.2)
to denote $\prod_{i=1}^ma[V_i]$ simply as $a[V_1|V_2|\cdots |V_n]$.
This notation is suggestive because the vertical bar can be interpreted
graphically as
a cut in the fermion propagator between the last gluon line of $V_i$
and the first gluon line of $V_{i+1}$. For a cut propagator,
instead of the usual factor
$(\sum_{j=1}^ix_{v_j}+i\epsilon)^{-1}$, we have
$-2\pi
i\delta(\sum_{j=1}^ix_{v_j})$. This notation is also convenient because
it makes (4.2) deceptively simple. It now reads
$a\{V_1;V_2:\cdots ;V_m\}=a[V_1|V_2|\cdots |V_m]$; we simply have to change the
semicolons to vertical bars.

Cut propagators are not limited to tree diagrams like Fig.~2. The
offshell gluons can be connected to other diagrams to
form a composite diagram that inherits the original cuts.
The cut diagrams so formed are similar to but different from the Cutkosky
cut diagrams. Similar because we have the same factors for the cut propagators.
Different because the cuts here
occur only on fermion lines whereas in a Cutkosky diagram they can occur on any
line. Moreover, via
(4.2), our cut diagram represents a sum of $M!/\prod_{j=1}^mn_j!$
(uncut) Feynman diagrams, with their real and imaginary parts fully included,
which is unlike the Cutkosky diagrams in which only the imaginary part
or the discontiuity is represented.

It is clear from (4.2) that the factorization formula can be thought of as
a sum rule, to represent sums of Feynman diagrams as cut diagrams.
As will be discussed in Appendix B, a cut diagram is easier to
compute than an uncut diagram. In this way not only it is unnecessary to
compute the individual diagrams first, the cut diagram representing the sum
is actually easier to compute than just one single Feynman diagram.

We shall now apply the factorization formula to compute sums of amplitudes
quoted in Sec.~2. We shall see that the $\ln s$ factor and the
$J_i$ functions that get cancelled out in the sum never once appear in the cut
diagrams.

In what follows we shall use the notation $\langle{B1}\rangle$ to denote the
spacetime part of $B1$ (without the color factor ${\bf G}_2$). Similar
notation will be used for the spacetime part of other diagrams as well.

\begin{figure}
\vskip -.3 cm
\centerline{\epsfxsize 3.0 truein \epsfbox {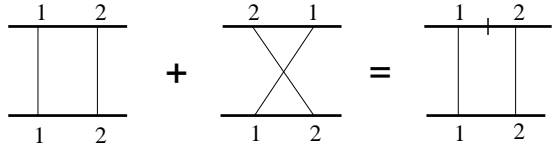}}
\nobreak
\vskip -0 cm\nobreak
\vskip .2 cm
\caption{ An illustration of summing Feynman diagrams to obtain cut 
diagrams.}
\end{figure}

In the 4th order, the ${\bf G}_2$ coefficient in ${\cal M}$ is given by
$\langle{B1}\rangle+\langle{B2}\rangle$. Using the eikonal formula
$a[12]+a[21]=a[1|2]$ on the upper fermion
line (see Fig.~3), $\langle{B1}\rangle+\langle{B2}\rangle$
is reduced to a cut diagram that can be easily calculated (see Appendix B
for all the calculations), yielding the correct result $g^2iI_2/2$
given in (2.1) and (2.4). Similary, the sum rule $a\{1;2;3\}=a[1|2|3]$ applied
to the six horizontal ladder diagrams in the 6th order, yields a
cut diagram (see Fig.~4) which gives the sum of $\langle{C15}\rangle$ to $\langle{C20}\rangle$ to be
$g^4I_3/6$. This is the correct coefficient of ${\bf G}_4$ given in (2.4).
Note that neither $\ln s$ nor $J_3$ ever appears.

\begin{figure}
\vskip -0 cm
\centerline{\epsfxsize 3 truein \epsfbox {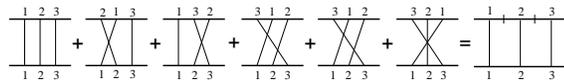}}
\nobreak
\vskip 0 cm\nobreak
\vskip .1 cm
\caption{Another sum rule.}
\end{figure}

\begin{figure}
\vskip -0 cm
\centerline{\epsfxsize 2.7 truein \epsfbox {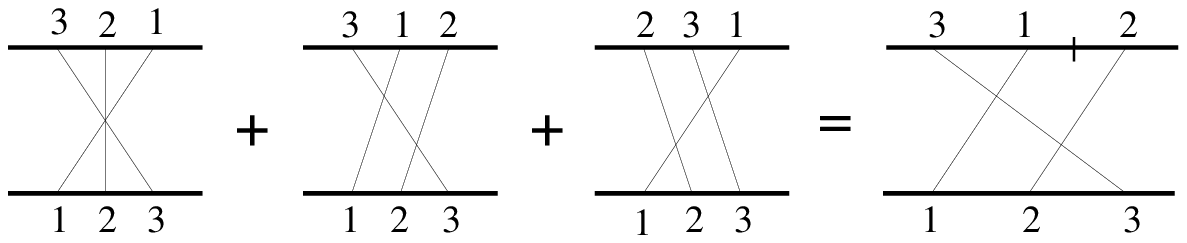}}
\nobreak
\vskip 0 cm\nobreak
\vskip .1 cm
\caption{Yet another sum rule.}
\end{figure}

\begin{figure}
\vskip -0 cm
\centerline{\epsfxsize 3.7 truein \epsfbox {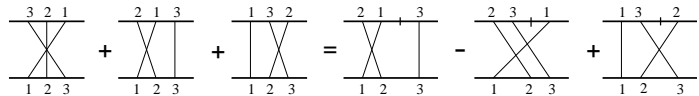}}
\nobreak
\vskip .1 cm\nobreak
\vskip .3 cm
\caption{An illustration of how sum of Feynman diagrams can be turned
into sum of cut diagrams.}
\end{figure}

Let us now look at the cancellation of $J_3(\Delta)$ in the ${\bf G}_2$ and
${\bf G}_3$ color factors, as discussed in point (5) of Sec.~2.
According to (2.1), the coefficient of  $-{\bf G}_3$ from diagrams C15 to C20 is given
by $\langle{C16}\rangle+\langle{C19}\rangle+\langle{C20}\rangle$. Using $a\{31;2\}=a[31|2]$ on the upper
fermion line as in Fig.~5, this sum is given by a cut diagram
which can be evaluated to be $-g^4i\ln s I_3/2\pi$. Again $J_3$ does not
appear in the cut diagram. Similarly, the coefficient of
$c{\bf G}_2$ from diagrams C15 to C20 is given by (2.1) to be
$3\langle{C16}\rangle+\langle{C17}\rangle+\langle{C18}\rangle+2\langle{C19}\rangle+2\langle{C20}\rangle$.
This is equal to twice the previous sum, plus $\langle{C16}\rangle+\langle{C17}\rangle+\langle{C18}\rangle$.
To use the sum rule, this requires an expression (see Fig.~6) for
$a[321]+a[213]+a[132]$. It is easy to see that this is equal to
$a\{21;3\}-a\{23;1\}+a\{13;2\}$, and by the factorization formula,
also equal to $a[21|3]-a[23|1]+a[13|2]$. There may seem
to be very little gained by replacing the sum of three terms by the sum
of another three terms, but this is not so. We are replacing the sum of
three uncut diagrams by the sum of three cut diagrams.
The computation of cut diagrams is much easier than the computation of
uncut diagrams. For one thing the complicated function $J_3$ that appears
on all the three uncut diagrams but disappears from their sum never appears
in any of the three cut diagrams. Evaluating the cut diagrams, the coefficient
of $c{\bf G}_2$ from C15 to C20 is $-g^4ic\ln s I_3/2\pi$, agreeing with the answer
given in (2.4).
\setcounter{equation}{0}
\section{Cut Diagrams}

The method introduced in the last section has a serious shortcoming.
It does not tell us
which cut diagrams to compute without detailed considerations
of the kind carried out there. In this section we discuss a remedy
for this shortcoming with the help of the
{\it multiple commutator formula} [12].

So far we have been treating Fig.~2 mostly as a QED or a scalar amplitude.
For QCD the non-abelian color matrices $t_a$ have to be incorporated. Instead
of (3.4) the amplitude is now
\begin{eqnarray}
A[v_1v_2\cdots v_n]=a[v_1v_2\cdots v_n]t[v_1v_2\cdots v_n]
\equiv a[V]t[V]\equiv A[V]\ ,
\end{eqnarray}
where $t[V]=t_{v_1}t_{v_2}\cdots t_{v_n}$. What we want is a formula for
the sum of the $n!$ permuted gluon orderings, ${\cal A}=\sum_{V\in S_n}A[V]$.
For QED, where we can take all $t_a=1$,
this is simply the eikonal formula, so what we need is the non-abelian
generalization of it. This is the multiple commutator formula
\begin{eqnarray}
{\cal A}\equiv\sum_{V\in S_n}a[V]t[V]=\sum_{V\in S_n}a[V_c]t[V_c']\
.
\end{eqnarray}
It expresses the sum of $a[V]t[V]$ in terms of sums over the corresponding cut
amplitude $a[V_c]t[V_c']$.
Compared to the eikonal formula this looks complicated; instead of
a single term on the right hand side we have now a sum over $n!$
terms. The complication is inevitable because we are attempting to sum up
amplitudes for {\it every} color. However,
we shall see that many of these terms are actually zero, and moreover,
the cut diagrams on the right are considerably simpler to evaluate than
the uncut diagrams on the left. Again delicate cancellations will largely be
avoided as before.

What remains to be described is what  the cut
diagram $V_c$ that corresponds to the Feynman diagram $V$ is, as well as what
the amplitudes $a[V_c]$ and $t[V_c']$ are.
Given a $V=[v_1v_2\cdots v_n]$,
start from the rightmost number $v_n$ and proceed leftward until
one comes to the first number less than $v_n$. Put a cut just to the right
of this number. Then start from this number and proceed leftward again
until one comes to the first number that is less than this number, and
another cut is put just to the right of this new minimum number. Continue
this way until the end and we have constructed the cut diagram $V_c$. For
example, for $n=2$, the 2 cut diagrams are $[12]_c=[1|2]$ and $[21]_c=[21]$.
For $n=3$,
the six cut diagrams are $[123]_c=[1|2|3]$, $[213]_c=[21|3]$, $[312]_c=[31|2]$,
$[132]_c=[1|32]$, $[231]_c=[231]$,
and $[321]_c=[321]$.

To each cut diagram we associate a spacetime cut amplitude
$a[V_c]$ as described in the last section. Namely, it is given by (3.4) except
the propagator at a cut is replaced by $-2\pi i\delta(\sum_jx_{v_j})$.

The complementary diagram $V'_c$ of a cut diagram $V_c$ is obtained as follows.
If a cut appears between two numbers in $V_c$, then
there will be no cut between the same two numbers in $V'_c$, and vice versa.
For $n=2$, the complementary cut diagrams are $[1|2]'=[12]$ and $[21]'=[2|1]$.
For $n=3$, the complementary cut diagrams are $[1|2|3]'=[123]$,
$[21|3]'=[2|13]$, $[31|2]'=[3|12]$, $[1|32]'=[13|2]$, $[231]'=[2|3|1]$, and
$[321]'=[3|2|1]$.

When no cut appears in $V'_c$ the color factor $t[V_c']$ is simply
$t[v_1v_2\cdots v_n]=t_{v_1}t_{v_2}\cdots  t_{v_n}$. If a cut appears between
$v_i$ and $v_{i+1}$, then the product $t_{v_i}t_{v_{i+1}}$ is replaced
by their commutators $[t_{v_i},t_{v_{i+1}}]$. If two or more consecutive
cuts appears, then the corresponding product of $t$'s is replaced by
multiple commutators. For example, $t[2|13]=[t_2,t_1]t_3$,
$t[2|3|1]=[t_2,[t_3,t_1]]$, and $t[4|3|2|15]=[t_4,[t_3,[t_2,t_1]]]t_5$.

\setcounter{equation}{0}
\section{Cut amplitudes to $O(g^6)$}

We shall compute the cut diagrams to $O(g^6)$ to
demonstrate the simplifications obtained therefrom.

\begin{figure}
\vskip -0 cm
\centerline{\epsfxsize 3.7 truein \epsfbox {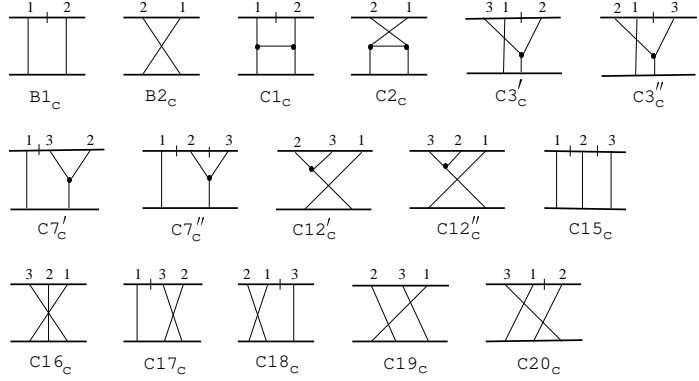}}
\nobreak
\vskip 0 cm\nobreak
\vskip .1 cm
\caption{cut diagrams up to the 6th order. }
\end{figure}

Using (5.2) on the upper fermion, we obtain
a set of cut diagrams as
shown in Fig.~7. This set is not unique as others can be obtained from a
different labelling of the gluon lines.

In the horizontal ladder diagrams B1, B2, C15--C20, we choose the planar
diagrams B1 and C15 to be the ones whose upper lines are completely cut.
This fixes the labelling [12] for B1 and  [123] for C15 as shown, and it
in turn determines how diagrams B2 and C16 to C20 are to be cut.
Using computational methods discussed in App.~A and B, these diagrams yield

\begin{eqnarray}
B1_c&=&+{1\over 2} g^2iI_2\!\cdot\!{\bf G}_2\nonumber\\
B2_c&=&+\beta(\ln s) I_2\!\cdot\!c{\bf G}_1\nonumber\\
C15_c&=&+g^4{1\over 6}I_3\!\cdot\!{\bf G}_4\nonumber\\
C16_c&=&-\beta^2(\ln s) 2J_3\!\cdot\!c^2{\bf G}_1\nonumber\\
C17_c&=&C18_c=0\nonumber\\
C19_c&=&0\nonumber\\
C20_c&=&-g^2\beta i(\ln s) I_3\!\cdot\!(c{\bf G}_2-{\bf G}_3)
\end{eqnarray}

Similarly, we choose to cut the line of the planar diagram C1 to obtain
\begin{eqnarray}
\overline{C1}_c&=-g^2\beta i(\ln s)\left[ {1\over 2}
\Delta^2I_2^2-J_2I_2\right] \!\cdot\!{\bf G}_3&\nonumber\\
\overline{C2}_c&=-\beta^2({\ln^2s})\left[ {1\over 2}
\Delta^2I_2^2-J_2I_2\right] \!\cdot\!c^2{\bf G}_1&
\end{eqnarray}

The twelve diagrams C3 to C20 can be divided into four groups of three, each
giving identical contributions, so it is necessary to consider only one
of these groups. The group of C3, C7, C12 have been chosen for that purpose.
There is a symmetry between gluon lines 2 and 3 so we may double this
group and consider it as a sum of six Feynman diagrams.
By applying the multiple commutator formula,
the six cut diagrams shown in Fig.~7 are obtained. Their values are
\begin{eqnarray}
C3_c&=&{1\over 2}(C3_c'+C3_c'')=-g^2\beta i(\ln s){1\over 4}J_2I_2\!\cdot\!{\bf G}_3\nonumber\\
C7_c&=&{1\over 2}(C7_c'+C7_c'')=0\nonumber\\
C12_c&=&{1\over 2}(C12_c'+C12_c'')=\beta^2(\ln^2\!s){1\over 4}J_2I_2\!\cdot\!(-c^2{\bf G}_1)\ .
\end{eqnarray}
The expressions in (6.1)--(6.3) should be compared with those of the uncut
diagrams, eq.~(2.1). It should also be compared with
the sum found in eq.~(2.4). Several points can be noted from these comparisons:
\begin{enumerate}
\item $\ln s$ factors that get cancelled in the sum of the Feynman
amplitude (see points (1) to (3) in Sec.~2) never even appear in (6.1)--(6.3).
Cancellation of this kind is automatically built into the cut diagram
formalism.
\item The tranverse function $J_3$ appears only in $C16_c$ in
eq.~(6.1). This expression survives the sum but can be ignored
compared to the contribution from $\overline{C2}_c$. In other words,
as opposed to the Feynman amplitude (2.1) where $J_3$ appears in many
places but most of them are cancelled out at the end (see points (4) and (5)
in Sec.~2), in the cut amplitude $J_3$ does not appear except when it
survives the sum.
\item The cut amplitude is not as successful in cancelling the
transverse function $J_2$ (point (6) of Sec.~2), although there is still
an imporvement here over (2.1) in that $J_2$ appears in fewer places. In fact,
it appears in $C3_c$ to $C20_c$ only when absolutely needed
to cancel its previous appearance in $\overline{C1}_c$ and $\overline{C2}_c$.
In order for $J_2$ to disappear completely it is necessary to combine
diagrams with triple and four gluon vertices together using the
Lipatov-Dickinson vertex \cite{EA}. The technique of cut diagrams by itself,
which deals mainly with the fermion lines, is not sufficient for that purpose.
\item Other than the $J_2$ complication mentioned above, the summands
of the final answer (2.4) appear directly in the cut amplitudes. In that sense
the cut amplitudes are as economical and as simple as they can ever be.
In particular, the Regge-like feature mentioned in point (3) of Sec.~2
is present already in individual cut diagrams.
\end{enumerate}

\section{Acknowledgments}
This research is supported in part by the by the Natural Science and 
Engineering Research Council of Canada, and the Fonds pour la 
Formation de Chercheurs et l'Aide \`a la Recherche of Qu\'ebec.
YJF acknowledges the support of the Carl Reinhart Foundation,
and OHR thanks the Ministry of Culture
and Higher Education of Iran for financial support.

\appendix
\setcounter{equation}{0}
\section{Color factors}
The method to compute the color factors graphically
\cite{CT} is briefly reviewed here. It is suitable for both cut and uncut
diagrams.

\begin{figure}
\vskip -0 cm
\centerline{\epsfxsize 2.7 truein \epsfbox {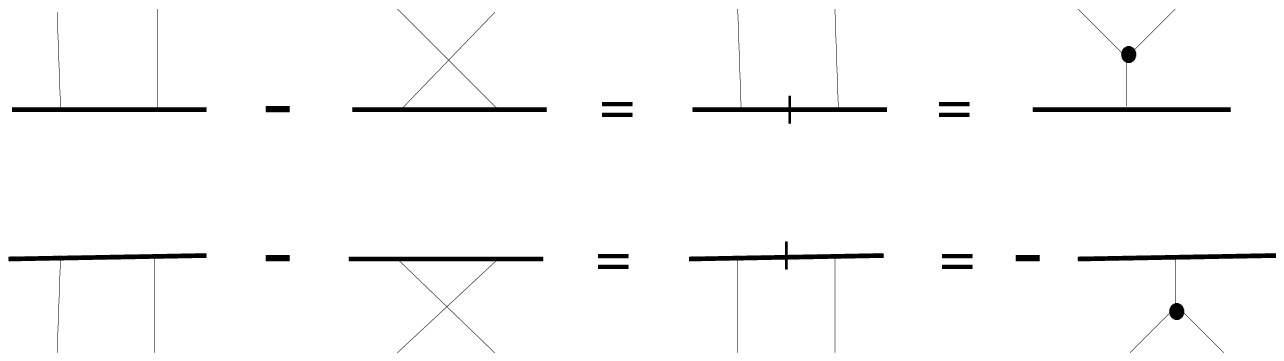}}
\nobreak
\vskip -0.5 cm\nobreak
\vskip .3 cm
\caption{Graphical represenation of the commutation relation of
color matrices. Thick lines are fermions and thin lines are gluons. The
color factor at a gluon-fermion vertex is $t_a$ of eq.~(2.3), and the 
color factor for a triple
gluon vertex whose color indices $a,b,c$ in the diagram are in clockwise order is $if_{abc}$. A
cut represents a commutator between two color matrices.}
\end{figure}

The basic tool is the graphical identities depicted in Figs.~8 and 9, which are
nothing but the commutation relation and identities in (2.3).
The commutation relation is valid for all representations of the color matrix
$t_a$, hence Fig.~8 remains true when the quark line is
replaced by a gluon line.

\begin{figure}
\vskip -0.3 cm
\centerline{\epsfxsize 3.2 truein \epsfbox {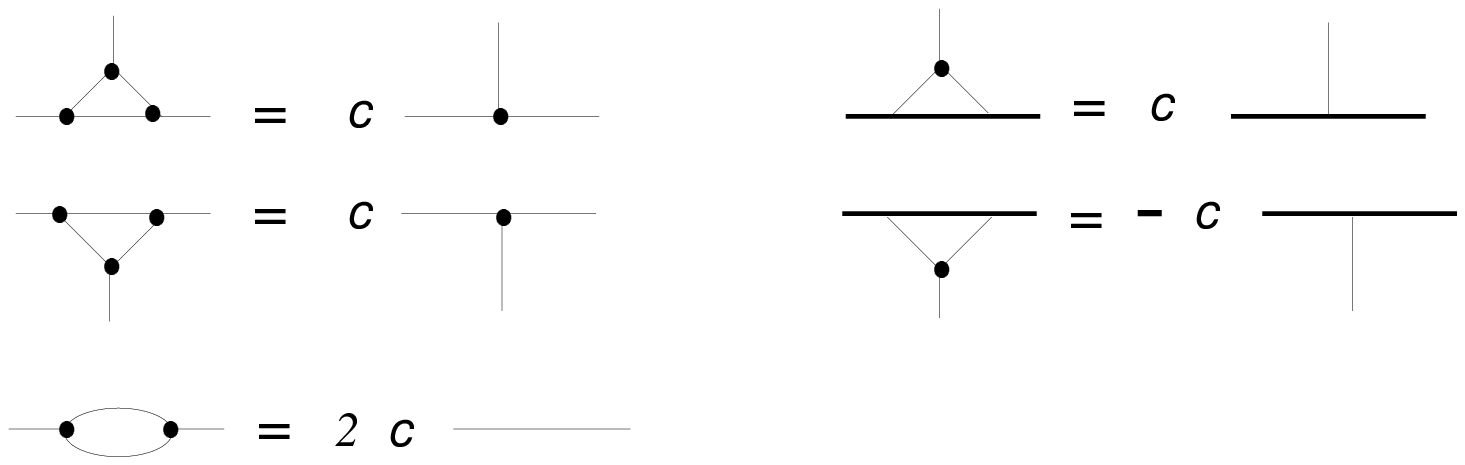}}
\nobreak
\vskip -.5 cm\nobreak
\vskip .1 cm
\caption{Graphical represenation of the last 
two identities in equation (2.3).  $c=N/2$ for $SU(N)$ color.}
\end{figure}

\begin{figure}
\vskip -1 cm
\centerline{\epsfxsize 3.2 truein \epsfbox {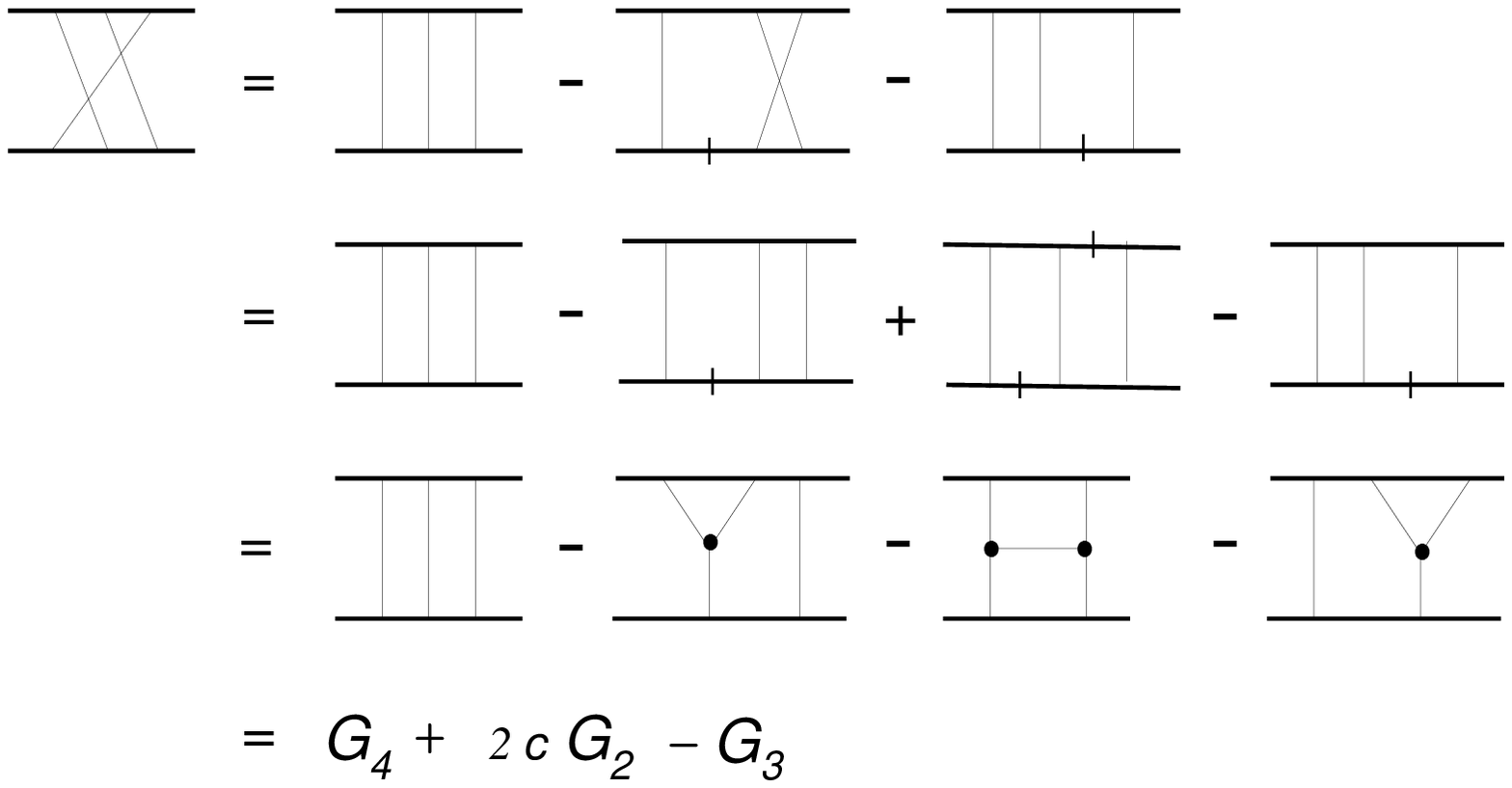}}
\nobreak
\vskip -.8 cm\nobreak
\vskip .1 cm
\caption{A sample calculation of the color factor of a Feynman diagram.
 }
\end{figure}

As an example, the computation of the color factor for the uncut diagram C19 in
eq.~(2.1) is carried out in Fig.~10, and the computation of the color factor
for the cut diagram  C19$_c$ in eq.~(6.1) is carried out in Fig.~11.
For the latter, we have to remember to use the {\it complementary} cut diagram
for the color factor, and not the cut diagram itself shown in Fig.~7.

\begin{figure}
\vskip 0 cm
\centerline{\epsfxsize 3.5 truein \epsfbox {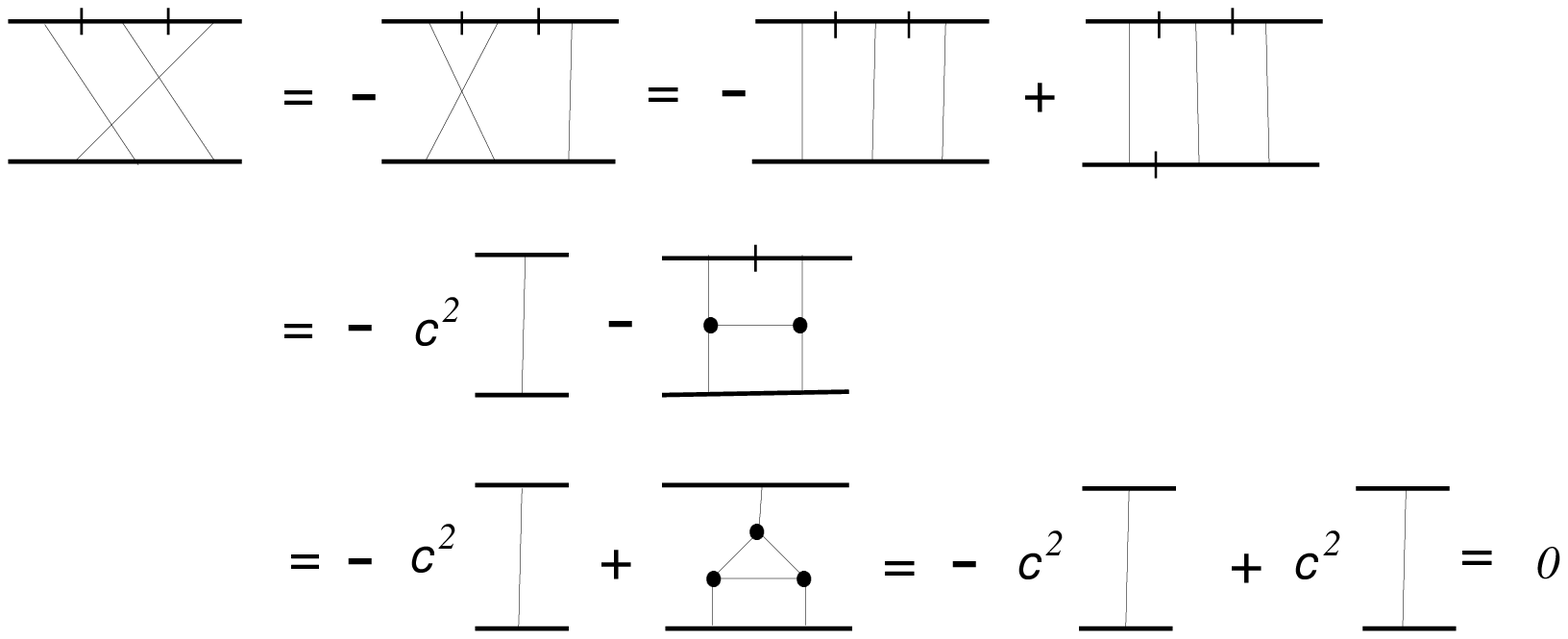}}
\nobreak
\vskip 0 cm\nobreak
\vskip .2 cm
\caption{A sample calculation of the color factor of a cut diagram.
 }
\end{figure}

The color factors in eqs.~(2.1), (6.1)--(6.3) can all be obtained in this way.

\setcounter{equation}{0}
\section{ Spacetime amplitudes}

High energy limit of a Feynman diagram can be computed either in the
Feynman-parameter
representation\cite{BW}, or directly in momentum space using
lightcone coordinates\cite{CT,NT,LN}. We follow the latter approach and review briefly
the main points. For a more detailed discussion see Ref.~\cite{CT}. It
serves  well to remember at this point that the bottom fermion carries mostly
the `$-$' momentum and the top fermion line carries mostly the `$+$'
momentum.

The procedure to follow for the computation is simple: (i) carry out the
`$+$' momentum integration
using residue calculus; (ii) then carry out the `$-$' momentum integration
to produce the $\ln s$ dependence. The transverse momentum integrations are never
explicitly carried out, which is why the answers are all expressed in terms
of functions like $I_n(\Delta)$ and $J_n(\Delta)$ given in eq.~(2.2).

\begin{figure}
\vskip -0 cm
\centerline{\epsfxsize 3.5 truein \epsfbox {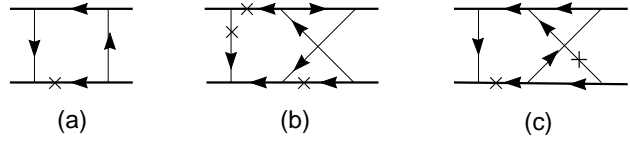}}
\nobreak
\vskip -0cm\nobreak
\vskip .1 cm
\caption{Flow diagrams. }
\end{figure}

Step (i) requires a knowledge of the location of poles for each
propagator, which in turn depends on the choice of loop momenta $k_a$.
Suppose $q=P+\sum_{a=1}^\ell
c_ak_a$ is the momentum flowing through a propagator, with $P$ being
some combination of external momenta and $c_a$ some integer
coefficients. The
propagator $D^{-1}=(q^2-m^2+i\epsilon)^{-1}=
(q_-q_+-q_\perp^2-m^2+i\epsilon)^{-1}$ has a pole in the
integration variable $k_{a+}$ if $c_a\not=0$. This pole is located in the upper
half plane if $c_aq_-$ is negative, and in the lower half plane if it
is positive. A way to keep track of the sign is to draw a
{\it flow diagram} for the `$-$' momentum,  with arrows indicating the
direction of the `$-$' flow. See Fig.~12 for some illustrative examples.
The arrows around a loop pointing one way (clockwise or
counter clockwise) have their
poles in one half plane, and those pointing the
other way have poles in the opposite half plane.
The arrows in the flow diagrams must obey momentum conservation, and one may
also assume they do not go around in closed loops.
Otherwise the poles will all be in the same half plane, and the integral is
zero if the contour is closed in the opposite
half plane. With these constraints we see in
Fig.~12 that a one-loop diagram allows only one flow path,
but for two and more loops there is bound to be
more than one flow diagrams because these constraints simply cannot fix
the direction of the `$-$' flow
on a boundary line of two loops.

Each flow diagram corresponds to a range of $k_{a-}$ variables.
By definition, the `$-$' variables along the direction of the flow
are always non-negative.

We shall always close integration contours in the lower
half planes, and indicate the poles so enclosed by a cross (x)
in the flow diagram. For a {\it scalar} diagram, the ${\cal T}$-matrix element is
equal to the product of $I$ propagators
$D^{-1}=(q^2-m^2+i\epsilon)^{-1}$, integrated over the $\ell$ loop momenta
$d^4k_a={1\over 2} dk_+dk_-d^2k_\perp$, with an extra numerical
factor $-[i/(2\pi)^4]^\ell$.
Each `$+$' integration produces a factor $-2\pi i$,
the ${\cal T}$-matrix is equal to
\begin{eqnarray}
{\cal T}=-\sum\int Dk_\perp\left(  \prod_{a=1}^\ell {dx_a\over 4\pi}\right){1\over
\prod_{i=1}^I D_i}\ ,
\end{eqnarray}
where $x_a=k_{a-}/\sqrt{s}$ is the scaled `$-$' momenta,
\begin{eqnarray}
Dk_\perp\equiv\prod_{a=1}^\ell {d^2k_{a\perp}\over (2\pi)^2}
\end{eqnarray}
is the measure for transverse momentum integration, and $D_i$ is either
the propagator evaluated at the x poles or the residue of the x pole
divided by $\sqrt{s}$. Each
summand in (B1) corresponds to a flow diagram with
a pole taken from the lower half plane (an x pole) of each
loop. A flow diagram may have more than one set of x poles, in which
case the sum is taken over all possible sets. For example, the one-loop
diagram B1 in Fig.~1 has only one flow path (Fig.~12(a)) and one
set of x poles . The two-loop diagram C17 of Fig.~1 has two flow paths,
Figs.~12(b) and 12(c). The first flow diagram has two sets of x poles,
and the latter flow diagram has one set of x poles. We shall see later
that in the leading log approximation, we may discard the $q_+$ dependence
on the upper fermion line and the accompanied poles, in which case
12(b) is also left with one set of x poles.

Equation (B1) is also valid for cut diagrams, provided the
propagator factor $D^{-1}=(q^2-m^2+i\epsilon)^{-1}$ of a cut line is replaced
by $-2\pi i\delta(q^2-m^2)$.

Let $q_{i-}=\sqrt{s}z_i,\ i=1,2,\cdots ,I$. Every $z_i$ is a linear combination
of $x_a$, and in the case of a propagator
along the bottom fermion line, $x_0\equiv 1$ is also involved.
The sign of each $z_i$ is fixed by the direction of arrows in the flow diagram.
If there are $\chi$ cut lines, then the last $\chi$ $x_a$ will be chosen
to be equal to $|z_i|$ of these cut lines.
The $I$ indices $i$ will now be divided into three sets, $a,b$ for indices
from 1 to $\ell$ labelling the internal lines with an x pole, $u$ for
indices labelling propagators on the top fermion line, and $m$ for the rest.
Then $q_{a+}=q_{a\perp}^2/\sqrt{s}z_a$, and $q_{m+}, q_{u+}$ can be expressed
as linear combinations of $q_{a+}$, {\it e.g.,}
\begin{eqnarray}
q_{m+}=\sum_ac_{ma}q_{a+}=z_m\sum_bc_{mb}q_{b\perp}^2/z_b\ 
\end{eqnarray}
Within the leading log approximation, one has
\begin{eqnarray}
D_a&=&z_a\nonumber\\
D_u&\simeq& sz_u\ (uncut)\ ,\quad D_u=-2\pi i\delta(sz_u)\ (cut)\nonumber\\
D_m&=&\sum_bc_{mb}{z_m\over z_b}q_{b\perp}^2-q_{m\perp}^2+i\epsilon\ .
\end{eqnarray}

The `$-$' momentum flows mainly along the bottom fermion line, with very little
seeping out to avoid a substantial mixing with the `$+$' momentum coming
from the top fermion line, for a finite mixture of these two at a propagator
would make it proportional to $s$ and therefore negligible.
This means the dominant contribution to ${\cal T}$ comes from regions where
all $x_a$ are small. Since $x_a=0$ for $a=\ell-\chi+1,\cdots ,\ell$, on account
of the $\delta$-functions of the cut lines,
the region of `$-$' integration can be roughly divided into regions
where the remaining $\ell'=\ell-\chi$ $x_a$'s are strongly ordered, and regions
where two or more of these $x_a$'s are of the same order of magnitude.
We shall label any one of the latter regions by $S$, and the former
regions by $R[12\cdots \ell']=\{1\gg x_1\gg x_2\gg\cdots \gg x_\ell'\ge a/s\}$ and its
permutations.

The $x$-dependence of $\prod_iD_i$ in $R[12\cdots \ell']$
is of the form $\prod_{a=1}^{\ell'}x_a^{-m_a}$, so the $x$-integral encountered
in (B1) is
\begin{eqnarray}
\int_{a\over s}{dx_{\ell'}\over x_{\ell'}^{m_{\ell'}}}\int_{x_{\ell'}}
{dx_{\ell'-1}\over x_{\ell'-1}^{m_{\ell'-1}}}\cdots \int_{x_2}{dx_1\over x_1^{m_1}}\
\sim {1\over s^M}(\ln s)^B\ ,
\end{eqnarray}
where $M=\sum_{a=1}^{\ell'}
(m_a-1)$ and $B$ is determined by how many times the sum $\sum_{a=1}^b(m_a-1)$
reaches zero by varying $b$ from 1 to $\ell'$. Clearly $B\le \ell'$,
and the only way for $B=\ell'$ is to have all $m_a=1$, in which case
we will call the
$\ln s$ dependence of ${\cal T}$ {\it saturated}. Otherwise it is said to be
{\it unsaturated}. For the uncut diagrams, $\chi=0$ and $\ell'=\ell$.
We see in (2.1) that all the diagrams except for the 6th
order horizontal ladder diagrams C15 to C20 are saturated. For the cut diagrams
in (6.1) to (6.3), only $C16_c$ is unsaturated, but this diagram does not
contribute to the final sum because of its subdominance.

The integral in an $S$ region is like one in an $R$ region with $\ell'$
reduced. For example, suppose $x_{\ell'-2}$ to $x_{\ell'}$ are roughly equal
but $x_1$ to $x_{\ell''}$ ($\ell''=\ell'-2$) are strongly
ordered
as in $R[12\cdots \ell'']=R''$. Then the volume element in the last three variables
in spherical coordinates is $r^2drd\Omega_2$, so the integration region is
effectively $R''$ but with $x_{\ell''}$ replaced by $r$. Since $\ell''<\ell$,
integrals in $S$ will never lead to saturation.

Cut diagrams are easier to compute than uncut Feynman
diagrams for three reasons. First and most trivial,
a cut line contains a $\delta(sz)$
which makes the corresponding `$-$' integration simpler to carry out.
Second and more importantly, the $\delta$-function
demands the absence of the `$-$' momentum across this line, so the cut line
is cut also in the sense of being an open electric circuit.
This generally reduces the number of
possible flow diagrams and makes the corresponding integral easier to saturate.
For example, each of the two-loop diagrams in Fig.~13
has only one flow path.
Thirdly, the flow pattern often leads to vanishing cut diagrams.
For example, both diagrams 13(c) and 13(d) are zero from the `$+$'
integration around the loop (12345) because
the `$-$' flows around that loop are all in the same direction.
This accounts for the equality $C17_c=C18_c=0$ in eq.~(6.1).

The cost of this simplicity is the presence of
{\it tricky diagrams}, which
are cut diagrams that are apparently logarithmic divergent at large $k_{a+}$.
This happens whenever there is a loop with only
one arrow present. Arrows on the top fermion line should not be counted
for this purpose because their approximate propagators $1/(sz+i\epsilon)$ no
longer carry any $k_{a+}$. Figs.~13(a)
and 13(b) are examples of such diagrams. This apparent divergence is produced
by the approximation of replacing the inverse propagators of an upper
line by $sz+i\epsilon$, thus losing some $q_+$ factors needed for convergence.
To regulate it we must replace the $\delta$-function of the cut line
by a smeared $\delta$-function, thereby allowing a small amount of `$-$'
momentum to flow through, and in the process restoring the lost $q_+$ factor.
If we do so to 13(a), the $k_+$ integration is no longer divergent, it will
have exactly the same flow path as in 12(a) and this is indicated in 13(a) by
the dotted arrows, so a pole can
be taken at the bottom fermion line as shown. The result is given by (B1) to be
$(g^4i/2s)I_2\!\cdot\!{1\over 2}$, the extra ${1\over 2}$ is there because the $x$ integration
is bounded by the flow path in 12(a) to be between 0 and 1, so only half
of the $\delta(sx)$ is integrated. When multiplied by the QED vertex
and normalization factor $(2s)^2/(2m)^2$, one obtains $B1_c$ in (6.1).

\begin{figure}
\vskip -0 cm
\centerline{\epsfxsize 3.2 truein \epsfbox {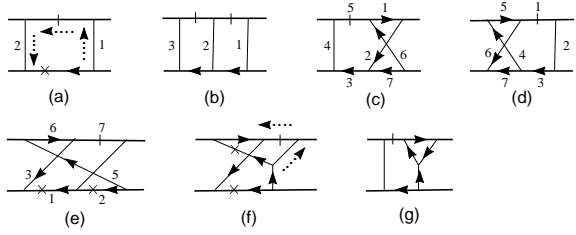}}
\nobreak
\vskip -0 cm\nobreak
\vskip .3 cm
\caption{Sample cut diagrams to illustrate their computation. }
\end{figure}

There is another way to compute Fig.~13(a). This is to recognize
the fact that the cut makes it symmetric in lines 1 and 2,
so we may replace the diagram by half the sum of it and its crossed diagram.
Using the sum rule (3.4),
$a[12]+a[21]=a[1|2]$, a cut can be produced at the bottom line, and the
resulting factor
$-2\pi i\delta(\sqrt{s}q_{1+})$ makes
the $q_{1+}$ integration convergent. See Fig.~14.
Moreover, the result of this
`$+$' integration is $-2\pi i$, exactly the same as if we were to do it by
residue calculus. The extra factor of ${1\over 2}$ obtained in the last paragrpah
now emerges because the double-cut diagram on the right of Fig.~14
is the sum of two diagrams.

\begin{figure}
\vskip -0 cm
\centerline{\epsfxsize 3.7 truein \epsfbox {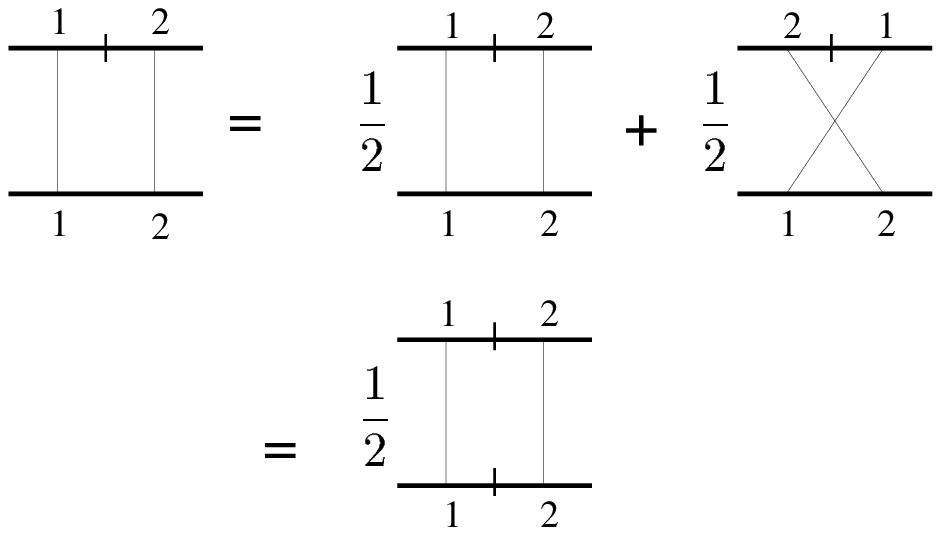}}
\nobreak
\vskip -.1 cm\nobreak
\vskip .1 cm
\caption{One-loop tricky diagram and its computation. }
\end{figure}

Now we come to the tricky diagram 13(b), which unlike the one-loop case is much
more difficult to compute by regulating the $\delta$-functions.
The reason is that there are now two cut lines, the relative magnitude of the
small `$-$' flows matters, and that produces once again two flow diagrams like
12(b) and 12(c). The poles are no longer at the bottom fermion line and
the computation is no simpler than the uncut diagrams. We must then compute
it by the second method. With the double cuts in 13(b), it is symmetric
in all the $q_{i-}$ variables. Insisting on this symmetry, 13(b) is equal
to $1/3!$ times the sum of 6 diagrams, obtained by permuting the bottom
gluon lines in all possible ways. From the factorization formula
$a\{1;2;3\}=a[1|2|3]$ used on the bottom line, the result is equal to
a diagram with all its fermion propagators cut. So 13(b) is given by
(B1) to be $g^6 (-2\pi i)^2 I_3/(4\pi s)^23!=-g^6I_3/24s^2$.
Incorporating the extra QED factor $-(2s)^3/(2m)^2$ for the 
${\cal T}$ matrix, we obtain the contribution of 
$C15_c$ to ${\cal M}\equiv(2m^2/g^2s){\cal T}$ shown in (6.1).

We conclude this appendix by discussing the remaining expressions in
eqs.~(6.1)--(6.3).
We shall use the notation $\langle{B2_c}\rangle$ to denote the spacetime contribution
to $B2_c$, etc. Then $\langle{B2_c}\rangle=\langle{B2}\rangle$, $\langle{C12_c'}\rangle=\langle{C_12_c''}\rangle=\langle{C12}\rangle$,
$\langle{C16_c}\rangle=\langle{C16}\rangle$, and $\langle{C19_c}\rangle=\langle{C19}\rangle$ can be obtained from (2.1).
The zero of $C19_c$ in (6.1) is due to the vanishing of the color factor
as shown in Fig.~11. Next, we compute $\langle{C3_c'}\rangle=\langle{C3_c''}\rangle$.
The flow path of this is shown in Fig.~13(f), with the dotted arrows
indicating the small regulating current which is allowed to flow only
in the direction shown. The calculation is identical to the uncut diagram \cite{CT},
except that the factor $-{1\over 2}{\ln^2s}$ is replaced by ${1\over 2}(\ln s)(-2\pi i)$ ($-2\pi i$
from the cut, $\ln s$ because only one uncut line is left on top, and ${1\over 2}$
because only half of the $\delta$-function is integrated). The result can then
be read off from (2.1) to be
\begin{eqnarray}
\langle{C3_c'}\rangle=\langle{C3_c''}\rangle=-g^2\beta i(\ln s){1\over 4}J_2I_2\ ,
\end{eqnarray}
which gives rise to the expression in (6.3).

Finally, we must show that $C7_c'=0$ and $C7_c''=0$. The former can be
seen from the flow path in Fig.~13(g), where the arrows around the small
triangle goes around in a closed loop. The latter is so because
the scalar diagram in Fig.~7 for $C7_c''$ is symmetrical in lines 2 and 3
but the triple gluon vertex is antisymmetrical in these two lines.


\begin{thebibliography}{99}
\bibitem[11]{note1}Note however
that for later convenience the femion arrows in Fig.~1 and all subsequent
figures are drawn from right to left, reflecting the order one writes matrices
as well as initial and final states on a piece of paper,
whereas in Ref.~\cite{CT} they are drawn in a
more conventional way from left to right.
\end{thebibliography}
\end{document}